\documentclass[12pt]{iopart}
\usepackage[utf8]{inputenc}
\usepackage{xspace}
\expandafter\let\csname equation*\endcsname=\relax
\expandafter\let\csname endequation*\endcsname=\relax
\usepackage{amsmath,amssymb,amsfonts,amsthm}

\catcode`,\active

\catcode`\,12

\catcode`,\active

\catcode`\,12

\newcommand{\Keywords}[1]{\par\noindent
{\footnotesize{\textbf{Keywords}\/}: #1}}
\newcommand{\nombrespacs}[1]{\par\noindent
{\small{PACS numbers\/}: #1}}

\newcommand{\pd}{\partial}

\newcommand{\under}[1]{_{#1}}
\numberwithin{equation}{section}
\begin{document}
\title[BI Algebra and a SI system on $S^2$]{The Bannai--Ito algebra and a superintegrable system with reflections on the 2-sphere}
\author{Vincent X. Genest}
\ead{genestvi@crm.umontreal.ca}
\address{Centre de recherches math\'ematiques, Universit\'e de Montr\'eal, Montr\'eal, Qu\'ebec, Canada, H3C 3J7}
\author{Luc Vinet}
\ead{vinetl@crm.umontreal.ca}
\address{Centre de recherches math\'ematiques, Universit\'e de Montr\'eal, Montr\'eal, Qu\'ebec, Canada, H3C 3J7}
\author{Alexei Zhedanov}
\ead{zhedanov@yahoo.com}
\address{Donetsk Institute for Physics and Technology, Donetsk 83114, Ukraine}
\begin{abstract}
A quantum superintegrable model with reflections on the 2-sphere is introduced. Its two algebraically independent constants of motion generate a central extension of the Bannai--Ito algebra. The Schr\"odinger equation separates in spherical coordinates and its exact solutions are presented. It is further observed that the Hamiltonian of the system arises in the addition of three representations of the $sl_{-1}(2)$ algebra (the dynamical algebra of the one-dimensional parabosonic oscillator). The contraction from the two-sphere to the Euclidean plane yields the Dunkl oscillator in two dimensions and its Schwinger-Dunkl symmetry algebra $sd(2)$.
\bigskip

\bigskip

\nombrespacs{02.30.Ik, 03.65.Fd}
\end{abstract}
\section{Introduction}
The class of superintegrable quantum systems is of particular interest as a laboratory for the study of symmetries, their algebraic description and their representations. A quantum system in $n$ dimensions with Hamiltonian $H$ is said to be maximally superintegrable if it possesses $2n-1$ algebraically independent constants of motion $c_i$ for $i=1,\ldots,2n-1$ commuting with $H$, $[H,c_i]=0$, where one of these constants is $H$. A system is of order $\ell$ if the maximum order in momenta of the constants of motion (other than $H$) is $\ell$. Empirically, superintegrable systems turn out to be exactly solvable.

The category of models that has been most analyzed is that of systems governed by scalar Hamiltonians of the form 
\begin{subequations}
\label{Type}
\begin{align}
H=\Delta+V,
\end{align}
where $\Delta$ is the standard Laplacian
\begin{align}
\Delta=\frac{1}{\sqrt{\mathrm{det}\,g}}\frac{\pd}{\pd x_i}\sqrt{\mathrm{det}\,g}\,g^{ij}\frac{\pd}{\pd x_{j}},
\end{align}
\end{subequations}
on spaces with metric $g_{ij}$ in coordinates $\{x_{i}\}$. In two dimensions, the second order superintegrable models of that type have been identified and classified \cite{Miller-2013-10}. As a matter of fact, they can all be obtained from the generic 3-parameter model on the two-sphere by limits (contractions) and specializations (see \cite{Kalnins-2013-05} for details). By observing that the Hamiltonian of the generic model on the two-sphere can be constructed through the addition of three $\mathfrak{su}(1,1)$ algebras, the constants of motion were identified in \cite{Genest-2013-tmp-2,Genest-2013-tmp-1} as the intermediate Casimir operators arising in the step-wise combination process. These were further shown to generate the Racah algebra which is the quadratic algebra with two independent generators that captures the bispectrality of the Racah polynomials sitting atop the discrete side of the Askey tableau of hypergeometric polynomials \cite{Koekoek-2010}. This identification of the symmetry algebra hence allows to associate the Racah polynomials to the generic 3-parameter superintegrable system on the 2-sphere. A threefold connection between the polynomials of the Askey scheme, the second-order superintegrable models and their symmetry algebras can further be achieved by performing on the Racah polynomials and the Racah algebras the contractions and specializations that lead from the generic model on the two-sphere to the other second-order superintegrable systems \cite{Kalnins-2013-05}.

The exploration of another category of superintegrable models has been undertaken recently: it bears on systems whose Hamiltonians involve reflections. Typically, the involutions arise in Dunkl operators \cite{Dunkl-1989-01} which are defined as follows in the special univariate case:
\begin{align}
\label{Dunkl-D}
\mathcal{D}_{x}^{\mu}=\pd_{x}+\frac{\mu}{x}(1-R_{x}),
\end{align}
where $\pd_{x}$ stands for the derivative with respect to $x$ and where $R_{x}$ is the reflection operator such that $R_xf(x)=f(-x)$. One of the simplest dynamical systems with reflections is the parabose oscillator in one dimension with Hamiltonian (see \cite{Kamefuchi-1982})
\begin{align*}
H=\left(\mathcal{D}_{x}^{\mu}\right)^2+x^2.
\end{align*}
Other one-dimensional models have been discussed (see for example \cite{Jafarov-2011-05,Plyushchay-2001,Post-2011-10}). Recall that Dunkl operators are most useful in the study of multivariate orthogonal polynomials associated to reflection groups \cite{Dunkl-2001} and that of symmetric functions \cite{Macdonald-1999} as well as in the analysis of exactly solvable quantum many-body systems of Calogero-Sutherland type (see for instance \cite{Vinet-1996}).

Interestingly, the theory of univariate orthogonal polynomials that are eigenfunctions of differential or difference operators of Dunkl type has also been the object of attention lately \cite{Genest-2013-09-02,Genest-2013-02-1,Tsujimoto-2012-03,Tsujimoto-2013-03-01,Vinet-2011-02,Vinet-2011-01,Vinet-2012-05}. These are now referred to as $-1$ polynomials and a scheme similar to the Askey one has emerged for them. Taking a place analogous to that of the Racah polynomials are the Bannai--Ito polynomials, which are sitting at the top of one side of the $-1$ tableau and which were introduced in a combinatorial context \cite{Bannai-1984}. The characteristic properties of these polynomials are encoded \cite{Genest-2013-12,Genest-2012,Tsujimoto-2012-03} in an algebra bearing the Bannai--Ito name which has 3 generators $L_1$, $L_2$ and $L_3$ verifying the following relations given in terms of anticommutators ($\{A,B\}=AB+BA$):
\begin{align}
\label{BI-Algebra}
\{L_1,L_2\}=L_3+\omega_3,\qquad \{L_2,L_3\}=L_1+\omega_1,\qquad \{L_3,L_1\}=L_2+\omega_2,
\end{align}
where $\omega_1$, $\omega_2$, $\omega_3$ are central. Introduced in \cite{Tsujimoto-2012-03}, the algebra \eqref{BI-Algebra} is the structure behind the bispectrality property of the Bannai--Ito polynomials. It corresponds to a $q\rightarrow -1$ limit of the Askey-Wilson algebra \cite{Zhedanov-1991-11},  which is the algebra behind the bispectrality property of the $q$-polynomials of the Askey scheme \cite{Koekoek-2010}; it has also been used in \cite{Vinet-2011} to study structure relations for $-1$ polynomials of the Bannai--Ito family. The special case with $\omega_1=\omega_2=\omega_3=0$ has been studied in \cite{Arik-2003, Gorodnii-1984} as an anticommutator version of the Lie algebra $\mathfrak{su}(2)$.

The examination of superintegrable systems with reflections has mostly focused so far on Dunkl oscillators in the plane \cite{Genest-2013-04,Genest-2013-09,Genest-2013-07} and in $\mathbb{R}^3$ \cite{Genest-2013-12-1}. These are formed out of combinations of one-dimensional parabose systems (with the inclusions of possible singular terms). They all are superintegrable and exactly solvable. In the ``isotropic'' case, the symmetry algebra denoted $sd(n)$ is a deformation of $\mathfrak{su}(n)$ with $n$ the number of dimensions. The Dunkl oscillators have proved to be showcases for $-1$ polynomials. An infinite family of higher order ($\ell>2$) superintegrable models with reflections has also been obtained with the help of the little $-1$ Jacobi polynomials \cite{Post-2011-12}. 

The purpose of this paper is to introduce and analyze an elegant superintegrable model with reflections on the two-sphere. The symmetry algebra will be seen to be a central extension of the Bannai--Ito algebra, a first physical occurrence as such of this algebra, as far as we know. This model-algebra pairing will present itself as the analog in the presence of reflections of the teaming of the Racah algebra with the so-called generic 3-parameter system on the two-sphere. It entails a relation \cite{Genest-2013-12} between Dunkl harmonic analysis on the 2-sphere and the representation theory of $sl_{-1}(2)$, a $q\rightarrow -1$ limit of the quantum algebra $\mathcal{U}_{q}(\mathfrak{sl}_2)$ that can be identified with the dynamical algebra of the parabose oscillator \cite{Tsujimoto-2011-10}.

The outline of the paper is as follows. In Section 2, the model is described, the constants of motion are exhibited and the invariance algebra they generate is identified as a central extension of the Bannai--Ito algebra. In section 3, the separated solutions of the model are given explicitly in two different spherical coordinate systems in terms of Jacobi polynomials and the symmetries responsible for the separation of variables are identified. In section 4, it will be shown how the model can be constructed from the addition of three $sl_{-1}(2)$ realizations and it will be seen that the constants of motion can be interpreted as Casimir operators arising in this Racah problem. The contraction from the two-sphere to the Euclidean plane will be examined in Section 5 and it will be shown how the Dunkl oscillator and its symmetry algebra are recovered in this limit. Some perspectives are offered in the conclusion.
\section{The model on $S^2$, superintegrability and symmetry algebra}
We shall begin by introducing the system on the 2-sphere that will be studied. Its symmetries will be given explicitly and the algebra they generate, a central extension of the Bannai--Ito algebra, will be presented.
\subsection{The model on $S^2$}
Let $s_1^2+s_2^2+s_3^2=1$ be the usual embedding of the unit two-sphere in the three-dimensional Euclidean space with coordinates $s_1$, $s_2$, $s_3$. Consider the model governed by the Hamiltonian 
\begin{align}
\label{Hamiltonian-S2}
\mathcal{H}=J_1^2+J_2^2+J_3^2+\frac{\mu_1}{s_1^2}(\mu_1-R_{1})+\frac{\mu_2}{s_2^2}(\mu_2-R_2)+\frac{\mu_3}{s_3^2}(\mu_3-R_3),
\end{align}
where the  $\mu_i$ are real parameters such that $\mu_i>-1/2$, a condition required for the normalizability of the wavefunctions (see section 3). The operators $J_i$ appearing in \eqref{Hamiltonian-S2} are the familiar angular momentum generators
\begin{align*}
J_1=\frac{1}{i}\left(s_2\pd_{s_3}-s_3\pd_{s_2}\right),\;
J_2=\frac{1}{i}\left(s_3 \pd_{s_1}-s_1 \pd_{s_3}\right),\;
J_3=\frac{1}{i}\left(s_1 \pd_{s_2}-s_2 \pd_{s_1}\right),
\end{align*}
that obey the $\mathfrak{so}(3)$ commutation relations
\begin{align*}
[J_1,J_2]=iJ_3,\quad [J_2,J_3]=iJ_1,\quad [J_3,J_1]=iJ_2.
\end{align*}
The operators $R_i$ in \eqref{Hamiltonian-S2} are the reflection operators with respect to the $s_i=0$ plane, i.e. $R_i f(s_i)=f(-s_i)$. Since these reflections are improper rotations, the Hamiltonian \eqref{Hamiltonian-S2} has a well defined action on functions defined on the unit sphere. In terms of the standard Laplacian operator $\Delta_{S^2}$ on the two-sphere \cite{Arfken-2012}, the Hamiltonian \eqref{Hamiltonian-S2} reads
\begin{align*}
\mathcal{H}=-\Delta_{S^2}+\frac{\mu_1}{s_1^2}(\mu_1-R_{1})+\frac{\mu_2}{s_2^2}(\mu_2-R_2)+\frac{\mu_3}{s_3^2}(\mu_3-R_3).
\end{align*}
\subsection{Superintegrability}
It is possible to exhibit two algebraically independent conserved quantities for the model described by the Hamiltonian \eqref{Hamiltonian-S2}. Let $L_1$ and $L_3$ be defined as follows:
\begin{subequations}
\label{Symmetries-1}
\begin{align}
L_1&=\left(i J_1+\mu_2\frac{s_3}{s_2}R_2-\mu_3 \frac{s_2}{s_3}R_3\right)R_2+\mu_2R_3+\mu_3 R_2+\frac{1}{2}R_2R_3,
\\
L_3 &=\left(i J_3+\mu_1\frac{s_2}{s_1}R_1-\mu_2 \frac{s_1}{s_2}R_2\right)R_1+\mu_1R_2+\mu_2 R_1+\frac{1}{2}R_1R_2.
\end{align}
\end{subequations}
A direct computation shows that one has 
\begin{align*}
[\mathcal{H}, L_1]=[\mathcal{H},L_3]=0,
\end{align*}
and hence $L_1$, $L_3$ are constants of the motion. Moreover, it can be checked that
\begin{align*}
[\mathcal{H},R_i]=0,\qquad i=1,2,3.
\end{align*}
and thus the reflection operators are also (discrete) symmetries of the system \eqref{Hamiltonian-S2}.

It is clear from \eqref{Symmetries-1} that $L_1$ and $L_3$ are algebraically independent from one another and hence it follows that the model with Hamiltonian \eqref{Hamiltonian-S2} on the two-sphere is maximally superintegrable. Since the constants of motion are of first order in the derivatives, the order of superintegrability is $\ell=1$. While this case is generally associated to geometrical symmetries and Lie invariance algebras for systems of the type \eqref{Type}, this is not so in the presence of reflections. In fact, as will be seen next, the invariance algebra is not a Lie algebra.
\subsection{Symmetry algebra}
To examine the algebra generated by the symmetries $L_1$ and $L_3$, it is convenient to introduce the operator $L_2$ defined as
\begin{align}
\label{Symmetries-2}
L_2=\left(-i J_2+\mu_1\frac{s_3}{s_1}R_{1}-\mu_3 \frac{s_1}{s_3}R_{3}\right)R_{1}R_2+\mu_1 R_3+\mu_3 R_1+\frac{1}{2}R_1R_3,
\end{align}
and the operator $C$ given by
\begin{align}
\label{Symmetries-3}
C=-L_1 R_2R_3-L_2 R_1R_3-L_3 R_1 R_2+\mu_1 R_1+\mu_2 R_2+\mu_3 R_3+\frac{1}{2}.
\end{align}
It is directly verified that both $L_2$ and $C$ commute with the Hamiltonian \eqref{Hamiltonian-S2}. Moreover, a straightforward calculation shows that $C$ also commutes with the symmetries
\begin{align*}
[C,L_i]=0,\qquad i=1,2,3.
\end{align*}
Furthermore, one can verify that the Hamiltonian of the system \eqref{Hamiltonian-S2} can be expressed in terms of $C$ as follows:
\begin{align*}
\mathcal{H}=C^2+C.
\end{align*}
Upon defining 
\begin{align*}
\mathcal{Q}=C R_1R_2R_3,
\end{align*}
which also commutes with the constants of motion $L_i$ and the Hamiltonian $\mathcal{H}$, it is verified that the following relations hold:
\begin{subequations}
\label{Invariance-Algebra}
\begin{align}
\{L_1,L_2\}&=L_3-2\mu_3\mathcal{Q}+2\mu_1\mu_2,
\\
\{L_2,L_3\}&=L_1-2\mu_1 \mathcal{Q}+2\mu_2\mu_3,
\\
\{L_3,L_1\}&=L_2-2\mu_2 \mathcal{Q}+2\mu_1\mu_3.
\end{align}
\end{subequations}
The invariance algebra \eqref{Invariance-Algebra} generated by the constants of motion $L_i$ of the system \eqref{Hamiltonian-S2} corresponds to a central extension of the Bannai--Ito algebra \eqref{BI-Algebra} where the central operator is $\mathcal{Q}$. In the realization \eqref{Symmetries-1}, \eqref{Symmetries-2}, \eqref{Symmetries-3}, the Casimir operator of the Bannai--Ito algebra, which has the expression \cite{Tsujimoto-2012-03}
$$
\mathbf{L}^2=L_1^2+L_2^2+L_3^2,
$$
is related to $C$ in the following way:
\begin{align*}
\mathbf{L}^2=C^2+\mu_1^2+\mu_2^2+\mu_3^2-1/4.
\end{align*}
Note that one has $C^2=\mathcal{Q}^2$ since $C$ commutes with $R_1R_2R_3$; further observe that the commutation relations \eqref{Invariance-Algebra} are invariant under any cyclic permutation of the pairs $(L_i,\mu_i)$, $i=1,2,3$. Since the reflections $R_i$ are also (discrete) symmetries of the Hamiltonian \eqref{Hamiltonian-S2}, their commutation relations with the other constants of motion $L_1$, $L_2$, $L_3$ can be included as part of the symmetry algebra. One finds that
\begin{align*}
\{L_i,R_j\}=R_{k}+2\mu_j R_{j}R_{k}+2\mu_k,\quad [L_i,R_i]=0,
\end{align*}
where $i\neq j\neq k$. The commutation relations involving $C$ and the reflections are
\begin{align*}
\{C,R_i\}=-2L_i R_{1}R_2R_{3}-R_i-2\mu_i,
\end{align*}
for $i=1,2,3$.
\section{Exact solution}
In this section, the exact solutions of the Schr\"odinger equation
\begin{align}
\label{SCH}
\mathcal{H}\Psi=\mathcal{E}\Psi,
\end{align}
associated to the Hamiltonian \eqref{Hamiltonian-S2} are obtained by separation of variables in two different spherical coordinate systems.
\subsection{Standard spherical coordinates}
In the usual spherical coordinates
\begin{align}
\label{Spherical-Standard}
s_1=\cos \phi \sin \theta,\quad s_2=\sin \phi \sin \theta,\quad s_3=\cos \theta,
\end{align}
the Hamiltonian \eqref{Hamiltonian-S2} takes the form
\begin{align}
\label{Hamiltonian-Spherical}
\mathcal{H}=\mathcal{F}_{\theta}+\frac{1}{\sin^2\theta}\,\mathcal{G}_{\phi},
\end{align}
where
\begin{subequations}
\label{Op-1}
\begin{flalign}
\mathcal{F}_{\theta}&=-\pd_{\theta}^2-\cot \theta\,\pd_{\theta}+\frac{\mu_3}{\cos^2\theta}(\mu_3-R_3),
\\
\label{Azimuthal-Op}
\mathcal{G}_{\phi}&=-\pd_{\phi}^2+\frac{\mu_1}{\cos^2\phi}(\mu_1-R_1)+\frac{\mu_2}{\sin^2\phi}(\mu_2-R_2),
\end{flalign}
\end{subequations}
and where the reflections have the actions
\begin{align*}
R_1f(\theta,\phi)=f(\theta,\pi-\phi),\quad R_2 f(\theta,\phi)=f(\theta,-\phi),\quad R_3 f(\theta,\phi)=f(\pi-\theta,\phi).
\end{align*}
It is clear from the expression \eqref{Hamiltonian-Spherical} that the Hamiltonian $\mathcal{H}$ separates in the spherical coordinates \eqref{Spherical-Standard}. Moreover, since $\mathcal{H}$ commutes with the three reflections $R_i$, they can all be diagonalized simultaneously. Upon taking $\Psi(\theta,\phi)=\Theta(\theta)\Phi(\phi)$ in \eqref{SCH}, one finds the system of ordinary equations
\begin{subequations}
\label{Eqs}
\begin{align}
\label{Zenithal}
\left[\mathcal{F}_{\theta}+\frac{m^2}{\sin^2\theta}-\mathcal{E}\right]\Theta(\theta)=0,
\\
\label{Azimuthal}
\left[\mathcal{G}_{\phi}-m^2\right]\Phi(\phi)=0,
\end{align}
\end{subequations}
where $m^2$ is the separation constant. The regular solutions to \eqref{Eqs} can be obtained from the results of \cite{Genest-2013-12-1}. Indeed, up to a gauge transformation with the function $G(s_1,s_2,s_3)=|s_1|^{\mu_1}||s_2|^{\mu_2}|s_3|^{\mu_3}$, the system \eqref{Eqs} is equivalent to the angular equations arising in the separation of variables in spherical coordinates of the Schr\"odinger equation for the three-dimensional Dunkl oscillator. 

Using this observation and the results of \cite{Genest-2013-12-1}, one finds that the azimuthal solutions have the following expression:
\begin{multline}
\label{Azimuthal-Sol}
\Phi_{n}^{(e_1,e_2)}(\phi)=\left(\frac{(n+\mu_1+\mu_2)\left(\frac{n-e_1-e_2}{2}\right)!\,\Gamma(\frac{n+e_1+e_2}{2}+\mu_1+\mu_2)}{2\,\Gamma(\frac{n+e_1-e_2}{2}+\mu_1+1/2)\Gamma(\frac{n+e_2-e_1}{2}+\mu_2+1/2)}\right)^{1/2}
\\
\,|\cos \phi|^{\mu_1}|\sin \phi|^{\mu_2}\,\cos^{e_1}\phi\sin^{e_2}\phi\;P_{(n-e_1-e_2)/2}^{(\mu_2-1/2+e_2,\mu_1-1/2+e_1)}(\cos 2\phi),
\end{multline}
where $P_{n}^{(\alpha,\beta)}(x)$ are the Jacobi polynomials \cite{Koekoek-2010}, $\Gamma(z)$ is the Gamma function \cite{Arfken-2012} and where $e_i\in \{0,1\}$. The azimuthal solutions \eqref{Azimuthal-Sol} satisfy the eigenvalue equations
\begin{align*}
R_1 \Phi_{n}^{(e_1,e_2)}(\phi)=(1-2e_1)\Phi_{n}^{(e_1,e_2)}(\phi),\qquad R_2 \Phi_{n}^{(e_1,e_2)}(\phi)=(1-2e_2)\Phi_{n}^{(e_1,e_2)}(\phi),
\end{align*}
with respect to the reflections. They obey the orthogonality relation
\begin{align*}
\int_{0}^{2\pi}\Phi_{n}^{(e_1,e_2)}(\phi)\,\Phi_{n'}^{(e_1',e_2')}(\phi)\;\mathrm{d}\phi=\delta_{nn'}\delta_{e_1e_1'}\delta_{e_2e_2'},
\end{align*}
as can be checked by comparing with the orthogonality relation satisfied by the Jacobi polynomials \cite{Koekoek-2010}. The separation constant here takes the value $m^2=(n+\mu_1+\mu_2)^2$. When $e_1+e_2=1$, $n$ is a positive odd integer, when $e_1+e_2=0$, $n$ is a non-negative even integer and when $e_1+e_2=2$, $n$ is a positive even integer. The regular solutions to the zenithal equation \eqref{Zenithal} are of the form \cite{Genest-2013-12-1}
\begin{multline}
\label{Zenithal-Sol}
\Theta_{n;N}^{(e_3)}(\theta)=\left(\frac{(N+\mu_1+\mu_2+\mu_3+1/2)(\frac{N-n-e_3}{2})!\Gamma(\frac{N+n+e_3}{2}+\mu_1+\mu_2+\mu_3+1/2)}{\Gamma(\frac{N+n-e_3}{2}+\mu_1+\mu_2+1)\Gamma(\frac{N-n+e_3}{2}+\mu_3+1/2)}\right)^{1/2}
\\
|\sin \theta|^{\mu_1+\mu_2}|\cos \theta|^{\mu_3}\sin^{n}\theta\cos^{e_3}\theta\;P_{(N-n-e_3)/2}^{(n+\mu_1+\mu_2,\mu_3-1/2+e_3)}(\cos 2\theta).
\end{multline}
with $e_3\in\{0,1\}$. The following eigenvalue equation holds:
\begin{align*}
R_3 \Theta_{n;N}^{(e_3)}(\theta)=(1-2e_3)\Theta_{n;N}^{(e_3)}(\theta).
\end{align*}
The energy $\mathcal{E}$ corresponding to the solution \eqref{Zenithal-Sol} is given by
\begin{align}
\label{Energy}
\mathcal{E}_{N}=(N+\mu_1+\mu_2+\mu_3)^2+(N+\mu_1+\mu_2+\mu_3),
\end{align}
where $N$ is a non-negative integer. The complete wavefunctions of the Hamiltonian \eqref{Hamiltonian-S2} on the 2-sphere with energy $\mathcal{E}_{N}$ given by \eqref{Energy} thus have the expression
\begin{align*}
\Psi_{n;N}^{(e_1,e_2,e_3)}(\theta,\phi)=\Theta_{n;N}^{(e_3)}(\theta)\Phi_{n}^{(e_1,e_2)}(\phi),
\end{align*}
where the zenithal and azimuthal parts are given by \eqref{Zenithal-Sol} and \eqref{Azimuthal-Sol}, respectively. By a direct counting of the admissible states (taking into account the fact that values of the quantum numbers $N$, $n$, $e_i$ yielding half-integer or negative values of $k$ in the Jacobi polynomials $P_{k}^{(\alpha,\beta)}(x)$ are not admissible), it is seen that the $\mathcal{E}_{N}$ energy eigenspace is $(2N+1)$-fold degenerate. Furthermore, one observes that
\begin{align*}
R_1R_2R_3 \Psi_{n;N}^{(e_1,e_2,e_3)}(\theta,\phi)=(-1)^{N}\Psi_{n;N}^{(e_1,e_2,e_3)}(\theta,\phi),
\end{align*}
and that the wavefunctions satisfy the orthogonality relation
\begin{align*}
\int_{0}^{\pi}\int_{0}^{2\pi}\Psi_{n;N}^{(e_1,e_2,e_3)}(\theta,\phi)\Psi_{n';N'}^{(e_1',e_2',e_3')}(\theta,\phi)\;\sin \theta\,\mathrm{d}\phi\,\mathrm{d}\theta=\delta_{nn'}\delta_{NN'}\delta_{e_1e_1'}\delta_{e_2e_2'}\delta_{e_3e_3'}.
\end{align*}
The symmetry operator responsible for the separation of variables in spherical coordinates is $L_3$. Indeed, a direct computation shows that upon defining $Z=L_3R_1R_2$ one has
\begin{align*}
Z^2-Z+\frac{1}{4}=\mathcal{G}_{\phi},
\end{align*}
where $\mathcal{G}_{\phi}$ is given by \eqref{Azimuthal-Op}.
\subsection{Alternative spherical coordinates}
The Schr\"odinger equation \eqref{SCH} associated to the Hamiltonian on the 2-sphere \eqref{Hamiltonian-S2} also separates in the alternative spherical coordinate system where the coordinates $s_1$, $s_2$, $s_3$ of the 2-sphere are parametrized as follows:
\begin{align}
\label{Alt}
s_1=\cos \vartheta,\quad s_2=\cos \varphi \sin \vartheta,\quad s_3=\sin \varphi\sin\vartheta.
\end{align}
In these coordinates the Hamiltonian \eqref{Hamiltonian-S2} takes the from
\begin{align*}
\mathcal{H}=\widetilde{\mathcal{F}}_{\vartheta}+\frac{1}{\sin^2\vartheta}\widetilde{\mathcal{G}}_{\varphi},
\end{align*}
where
\begin{subequations}
\label{Op-2}
\begin{flalign}
\widetilde{\mathcal{F}}_{\vartheta}&=-\pd_{\vartheta}^2-\cot \vartheta\,\pd_{\vartheta}+\frac{\mu_1}{\cos^2\vartheta}(\mu_1-R_1),
\\
\widetilde{\mathcal{G}}_{\varphi}&=-\pd_{\varphi}^2+\frac{\mu_2}{\cos^2\varphi}(\mu_2-R_2)+\frac{\mu_3}{\sin^2\varphi}(\mu_3-R_3),
\end{flalign}
\end{subequations}
and where the reflections have the actions
\begin{align}
R_1f(\vartheta,\varphi)=f(\pi-\vartheta,\varphi),\quad R_2 f(\vartheta,\varphi)=f(\vartheta,\pi-\varphi),\quad R_3 f(\vartheta,\varphi)=f(\vartheta,-\varphi).
\end{align}
Upon comparing \eqref{Op-2} with \eqref{Op-1} it is clear that the solutions to the Schr\"odinger equation \eqref{SCH} in the alternative coordinate system \eqref{Alt} have the expression
\begin{align*}
\Psi_{n;N}^{(e_1,e_2,e_3)}(\vartheta,\varphi)=\pi \,\Theta_{n;N}^{(e_1)}(\vartheta)\Phi_{n}^{(e_2,e_3)}(\varphi),
\end{align*}
where $\pi=(123)$ indicates the permutation applied to the parameters $(\mu_1,\mu_2,\mu_3)$. The symmetry associated to the separation of variables in the alternative coordinate system \eqref{Alt} is $L_1$ since upon taking $Y=L_1R_2R_3$, one finds that
\begin{align*}
Y^2-Y+\frac{1}{4}=\widetilde{\mathcal{G}}_{\varphi}.
\end{align*}
As illustrated above, the origin of the two independent constants of motion $L_3$ and $L_1$ of the model \eqref{Hamiltonian-S2} on the two-sphere can be traced back to the multiseparability of the Schr\"odinger equation in the usual and alternative spherical coordinates, respectively. This situation is analogous to the one arising in the analysis of the generic three-parameter system on the 2-sphere (without reflections) for which the symmetries generating the Racah algebra are associated to the separation of variables in different spherical coordinate systems \cite{Genest-2013-tmp-2}. In this case, the expansion coefficients coefficients between the separated wavefunctions in the coordinate systems \eqref{Spherical-Standard} and  \eqref{Alt} coincide with the $6j$ symbols of $\mathfrak{su}(1,1)$.
\section{Connection with $\mathbf{sl_{-1}(2)}$}
In this section, it is shown how of the Hamiltonian \eqref{Hamiltonian-S2} of the model on the 2-sphere arises in the combination of three realizations of the $sl_{-1}(2)$ algebra, which we loosely refer to as the Racah problem for $sl_{-1}(2)$.
\subsection{$sl_{-1}(2)$ algebra}
The $sl_{-1}(2)$ algebra was introduced in \cite{Tsujimoto-2011-10} as a $q\rightarrow -1$ limit of the quantum algebra $\mathcal{U}_{q}(\mathfrak{sl}_2)$. It has three three generators $A_{\pm}$, $A_0$ and one involution $P$ and is defined by the relations
\begin{align}
\label{sl}
[A_0,A_{\pm}]=\pm A_{\pm},\; [A_0,P]=0,\;\{A_{+},A_{-}\}=2A_0,\;\{A_{\pm},P\}=0,\;P^2=1.
\end{align}
The Casimir operator of $sl_{-1}(2)$, which commutes with all generators, is given by
\begin{align*}
Q=A_{+}A_{-}P-A_0P+P/2.
\end{align*}
Let $A_0^{(i)}$, $A_{\pm}^{(i)}$ and $P^{(i)}$, $i=1,2,3$, denote three mutually commuting sets of $sl_{-1}(2)$ generators. Using the coproduct of $sl_{-1}(2)$ (see \cite{Tsujimoto-2011-10}), the three sets can be combined to produce a fourth set of generators satisfying the relations \eqref{sl}. The elements of this fourth set $\{\mathcal{A}_0,\mathcal{A}_{\pm},\mathcal{P}\}$ are defined by
\begin{subequations}
\label{Total-Algebra}
\begin{align}
\mathcal{A}_0&=A_{0}^{(1)}+A_{0}^{(2)}+A_0^{(3)},
\\
\mathcal{A}_{\pm}&=A_{\pm}^{(1)}P^{(2)}P^{(3)}+A_{\pm}^{(2)}P^{(3)}+A_{\pm}^{(3)},
\\
\mathcal{P}&=P^{(1)}P^{(2)}P^{(3)}.
\end{align}
\end{subequations}
It is easily verified that the operators \eqref{Total-Algebra} indeed satisfy the defining relations \eqref{sl} of $sl_{-1}(2)$. In the combining of these independent sets of $sl_{-1}(2)$ generators, three types of Casimir operators should be distinguished. The initial Casimir operators $Q^{(i)}$
\begin{align*}
Q^{(i)}=A_{+}^{(i)}A_{-}^{(i)}P^{(i)}-A_0^{(i)}P^{(i)}+P^{(i)}/2,\quad i=1,2,3,
\end{align*}
which are attached to each independent set of $sl_{-1}(2)$ generators.
The two intermediate Casimir operators $\mathcal{Q}^{(ij)}$
\begin{align*}
Q^{(ij)}=(A_{-}^{(i)}A_{+}^{(j)}-A_{+}^{(i)}A_{-}^{(j)})P^{(i)}+Q^{(i)}P^{(j)}+Q^{(j)}P^{(i)}-P^{(i)}P^{(j)}/2,
\end{align*}
 for $(ij)=(12),\,(23)$ which are associated to the step-wise combination process. The total Casimir operator $\mathcal{Q}$ associated to the fourth set
\begin{align*}
\mathcal{Q}=\mathcal{A}_{+}\mathcal{A}_{-}\mathcal{P}-\mathcal{A}_0\mathcal{P}+\mathcal{P}/2,
\end{align*}
which has the expression
\begin{align*}
\mathcal{Q}=(A_{-}^{(1)}A_{+}^{(3)}-A_{+}^{(1)}A_{-}^{(3)})P^{(1)}-Q^{(2)}P^{(1)}P^{(3)}+Q^{(12)}P^{(3)}+Q^{(23)}P^{(1)}.
\end{align*}
The intermediate Casimir operators commute with both the initial and the total Casimir operators and with $\mathcal{P}$, but do not commute amongst themselves. As a matter of fact, it was established in \cite{Genest-2012} that on representations spaces where the total Casimir operator $\mathcal{Q}$ is diagonal, the intermediate Casimir operators $Q^{(ij)}$ generate the Bannai--Ito algebra.
\subsection{Differential/Difference realization and the model on the 2-sphere}
The connection between the combination of three $sl_{-1}(2)$ algebras and the superintegrable system on the two-sphere defined by \eqref{Hamiltonian-S2} can now be established. Consider the following differential/difference realization of $sl_{-1}(2)$:
\begin{subequations}
\label{Dunkl-Realization}
\begin{align}
A_{\pm}^{(i)}&=\frac{1}{\sqrt{2}}\left[s_i\mp \pd_{s_i}\pm \frac{\mu_i}{s_i}R_i\right],
\\
A_0^{(i)}&=\frac{1}{2}\left[-\pd_{s_i}^2+s_i^2+\frac{\mu_i}{s_i^2}(\mu_i-R_i)\right],
\\
P^{(i)}&=R_i.
\end{align}
\end{subequations}
Note that with respect to the uniform measure on the real line, $A_0^{(i)}$ is Hermitian and $A^{(i)}_{\pm}$ are Hermitian conjugates. Up to a gauge transformation of the generators
$$
z\rightarrow G(s_i)^{-1}zG(s_i),\qquad z\in\{A_{\pm}^{(i)},A_{0}^{(i)},P^{(i)}\},
$$
with gauge function $G(s_i)=|s_i|^{\mu_i}$, the realization \eqref{Dunkl-Realization} is equivalent to the realization of $sl_{-1}(2)$ arising in the one-dimensional parabose oscillator \cite{Genest-2013-04}. Using \eqref{Dunkl-Realization}, the initial Casimir operators $Q^{(i)}$ are seen to have the action
\begin{align*}
Q^{(i)}f(s_i)=-\mu_i f(s_i).
\end{align*}
and the intermediate Casimir operators take the form
\begin{align}
\label{Inter}
\begin{aligned}
Q^{(12)}=\left[(s_1\pd_{s_2}-s_2 \pd_{s_1})+\mu_1\frac{s_2}{s_1}R_1-\mu_2 \frac{s_1}{s_2}R_2\right]R_1+\mu_1R_2+\mu_2 R_1+\frac{1}{2}R_1R_2,
\\
Q^{(23)}=\left[(s_2\pd_{s_3}-s_3 \pd_{s_2})+\mu_2\frac{s_3}{s_2}R_2-\mu_3 \frac{s_2}{s_3}R_3\right]R_2+\mu_2R_3+\mu_3 R_2+\frac{1}{2}R_2R_3.
\end{aligned}
\end{align}
Furthermore, an explicit computation shows that upon defining $\Omega=\mathcal{Q}\mathcal{P}$, one finds
\begin{multline}
\label{Full}
\Omega^2+\Omega=
\\
J_1^2+J_2^2+J_3^2+(s_1^2+s_2^2+s_3^2)\left(\frac{\mu_1}{s_1^2}(\mu_1-R_1)+\frac{\mu_2}{s_2^2}(\mu_2-R_2)+\frac{\mu_3}{s_3^2}(\mu_3-R_3)\right).
\end{multline}
Upon comparing the expressions \eqref{Inter} for the intermediate Casimir operators with the formulas \eqref{Symmetries-1} for the constants of motion, it is seen that
\begin{align*}
Q^{(12)}=-L_3,\qquad Q^{(23)}=-L_1,
\end{align*}
and thus that the intermediate Casimir coincide with the constants of motion. Upon comparing \eqref{Full} with the Hamiltonian \eqref{Hamiltonian-S2}, it is also seen that
\begin{align*}
\Omega^2+\Omega=\mathcal{H},
\end{align*}
given the condition $s_1^2+s_2^2+s_3^2=1$. This condition can be ensured in general. Indeed,  one checks that
\begin{align*}
X^2=\frac{1}{2}(\mathcal{A}_{+}+\mathcal{A}_{-})^2=s_1^2+s_2^2+s_3^2.
\end{align*}
Since $X^2$ commutes with $\Omega$ and all the intermediate Casimir operators, it is central in the invariance algebra \eqref{Invariance-Algebra} and can thus be treated as a constant. Hence one can take $X^2=1$ without loss of generality and complete the identification of the quadratic combination $\Omega^2+\Omega$ with the Hamiltonian $\mathcal{H}$.

The analysis of the model on the 2-sphere defined by the Hamiltonian \eqref{Hamiltonian-S2}  is thus related to the combination of three independent realizations of the $sl_{-1}(2)$ algebra. The constants of motion of the system correspond to the intermediate Casimir operators arising in this combination and the Hamiltonian is related to a quadratic combination of the total Casimir operator (times $\mathcal{P}$). It is worth pointing out that in \cite{Genest-2013-tmp-2,Genest-2013-tmp-1} the Hamiltonian of the three-parameter model on the two-sphere (without reflections) was directly identified to the total Casimir operator in the combination of three $\mathfrak{su}(1,1)$ realizations. Here the total Casimir operator $\mathcal{Q}$ (or equivalently $\Omega$) is of \emph{first} order in derivatives and hence a quadratic combination must me taken to recover the Hamiltonian \eqref{Hamiltonian-S2} of the model. As a remark, let us note that the relation \eqref{Full} is reminiscent of chiral supersymmetry. Indeed, if one defines a new Hamiltonian by $H=\mathcal{H}+1/4$, then $H=\frac{1}{2}\{Q,Q\}$ where $Q=\Omega+1/2$.  In this picture, $Q=\Omega+1/2$ can be interpreted as a chiral supercharge for $H$. See also \cite{Genest-2013-10} for related considerations.
\section{Superintegrable model in the plane from contraction}
The results obtained so far are analogous to those of \cite{Genest-2013-tmp-2,Genest-2013-tmp-1} where the analysis of the generic 3-parameter system on the two-sphere was cast in the framework of the Racah problem for the $\mathfrak{su}(1,1)$ algebra. Given that the model \eqref{Hamiltonian-S2} is the analogue with reflections of the generic 3-parameter system on the 2-sphere and since all second-order superintegrable systems can be obtained from the latter \cite{Kalnins-2013-05}, it is natural to ask whether contractions of \eqref{Hamiltonian-S2} lead to other superintegrable systems with reflections. The answer to that question is in the positive. As an example, we describe in this section how the Dunkl oscillator model in the plane and its conserved quantities can be obtained from a contraction of the system \eqref{Hamiltonian-S2} on the two-sphere and its symmetries.

\subsection{Contraction of the Hamiltonian}
Consider the Hamiltonian
\begin{align}
\label{H2}
\mathcal{H}=J_1^2+J_2^2+J_3^2+(s_1^2+s_2^2+s_3^2)\left(\frac{\mu_1^2-\mu_1 R_1}{s_1^2}+\frac{\mu_2^2-\mu_2R_2}{s_2^2}+\frac{\mu_3^2-\mu_3R_3}{s_3^2}\right),
\end{align}
which is equivalent to \eqref{Hamiltonian-S2} in view of the results of section 4. The two-sphere $s_1^2+s_2^2+s_3^3=r^2$ of radius $r$ can be contracted to the Euclidean plane with coordinates $x_1$, $x_2$ by taking the limit as $r\rightarrow \infty$ in
\begin{align}
\label{Variables}
x_1=r\frac{s_1}{s_3},\quad x_2=r\frac{s_2}{s_3},\quad s_3^2=r^2-s_1^2-s_2^2,
\end{align}
Changing the variables in \eqref{H2} according to the prescription \eqref{Variables} and defining $\mu_3=\widehat{\mu}_3r^2$, a direct computation shows that
\begin{align}
\label{Dompe-2}
\begin{aligned}
\widetilde{\mathcal{H}}\equiv\lim_{r\rightarrow \infty}\frac{1}{r^2}&\Big(\mathcal{H}-\mu_3^2+\mu_3 R_3\Big)=
\\
&-\pd_{x_1}^2-\pd_{x_2}^2+\frac{\mu_1}{x_1^2}(\mu_1-R_1)+\frac{\mu_2}{x_2^2}(\mu_2-R_2)+\widehat{\mu}_3^2(x_1^2+x_2^2).
\end{aligned}
\end{align}
The Hamiltonian $\widetilde{\mathcal{H}}$ corresponds, up to a gauge transformation, to the Hamiltonian of the Dunkl oscillator model in the plane. Indeed, taking $\widetilde{\mathcal{H}}\rightarrow |x_1|^{-\mu_1}|x_2|^{-\mu_2}\widetilde{\mathcal{H}}|x_2|^{\mu_2}|x_1|^{\mu_1}$, one finds that
\begin{align}
\label{Dompe}
\widetilde{\mathcal{H}}\rightarrow -\left[(\mathcal{D}_{x_1}^{\mu_1})^2+(\mathcal{D}_{x_2}^{\mu_2})^2\right]+\widehat{\mu}_3^2(x_1^2+x_2^2),
\end{align}
where $\mathcal{D}_x^{\mu}$ is the Dunkl derivative \eqref{Dunkl-D}. Taking $\widehat{\mu}_3=1$, the Hamiltonian \eqref{Dompe} coincides with that of the model examined in \cite{Genest-2013-04}.
\subsection{Contraction of the constants of motion}
The conserved quantities of the Dunkl oscillator in the plane can be recovered by a contraction of the symmetry operators of the Hamiltonian \eqref{H2}. Since the reflections commute with \eqref{H2}, one can consider the following form of the constants of motion:
\begin{align*}
\widetilde{L}_1&=\frac{1}{i}\left[(s_2\pd_{s_3}-s_3 \pd_{s_2})+\mu_2 \frac{s_3}{s_2}R_2-\mu_3 \frac{s_2}{s_3}R_3\right],
\\
\widetilde{L}_2&=\frac{1}{i}\left[(s_3 \pd_{s_1}-s_1 \pd_{s_3})+\mu_3 \frac{s_1}{s_3}R_3-\mu_1 \frac{s_3}{s_1}R_1\right],
\\
\widetilde{L}_3 &= \frac{1}{i}\left[(s_1 \pd_{s_2}-s_2 \pd_{s_1})+\mu_1\frac{s_2}{s_1}R_1-\mu_2 \frac{s_1}{s_2}R_2\right],
\end{align*}
in lieu of the symmetries $L_1$, $L_2$, $L_3$ in \eqref{Symmetries-1} and \eqref{Symmetries-2}. The operators $\widetilde{L}_i$ commute with \eqref{H2} and satisfy the relations
\begin{align*}
[\widetilde{L}_1,\widetilde{L}_2]=i \widetilde{L}_3(1+2\mu_3 R_3),
\\
[\widetilde{L}_2,\widetilde{L}_3]=i \widetilde{L}_1(1+2\mu_1 R_1),
\\
[\widetilde{L}_3,\widetilde{L}_1]=i \widetilde{L}_2(1+2\mu_2 R_2).
\end{align*}
The commutation relations with the reflections are given by
\begin{align*}
\{\widetilde{L}_i, R_j\}=[\widetilde{L}_i, R_i]=0,
\end{align*}
where $i\neq j$. The contraction of $\widetilde{L}_3$ directly yields a conserved quantity for \eqref{Dompe-2}. Indeed, one finds using \eqref{Variables}
\begin{align}
\label{J2}
\mathcal{J}_2\equiv\lim_{r\rightarrow \infty}\widetilde{L}_3=\frac{1}{i}\left[(x_1\pd_{x_2}-x_2 \pd_{x_1})+\mu_1\frac{x_2}{x_1}R_1-\mu_2 \frac{x_1}{x_2}R_2\right],
\end{align}
which commute with $\widetilde{\mathcal{H}}$ given by \eqref{Dompe-2}.
The contraction of the symmetries $\widetilde{L}_1$, $\widetilde{L}_2$ cannot lead to constants of motion for the resulting Hamiltonian \eqref{Dompe-2} since these two operators anticommute with the term involving the reflection operator $R_3$ which is added before the $r\rightarrow \infty$ limit is taken (see \eqref{Dompe-2}). However, the contraction of $\widetilde{L}_1^2$ and $\widetilde{L}_2^2$, which commute with $R_3$, will yield symmetries of the contracted Hamiltonian \eqref{Dompe-2}. Computing the squares of $\widetilde{L}_1$, $\widetilde{L}_2$ and using \eqref{Variables}, one finds
\begin{align*}
\widetilde{\mathcal{H}}_2\equiv \lim_{r\rightarrow \infty}\frac{1}{r^2}\left(\widetilde{L}_1^2+\mu_3 R_3+2\mu_2\mu_3 R_2 R_3\right)=-\pd_{x_2}^2+\widehat{\mu}_3^2 x_2^2+\frac{\mu_2}{x_2^2}(\mu_2-R_2),
\\
\widetilde{\mathcal{H}}_1\equiv\lim_{r\rightarrow \infty}\frac{1}{r^2}\left(\widetilde{L}_2^2+\mu_3 R_3+2\mu_1\mu_3 R_1 R_3\right)=-\pd_{x_1}^2+\widehat{\mu}_3^2 x_1^2+\frac{\mu_1}{x_1^2}(\mu_1-R_1).
\end{align*}
It is clear that $\widetilde{\mathcal{H}}=\widetilde{\mathcal{H}}_1+\widetilde{\mathcal{H}}_2$ and hence one can define
\begin{align}
\label{J1}
\mathcal{J}_1=\widetilde{\mathcal{H}}_1-\widetilde{\mathcal{H}}_2,
\end{align}
as a second constant of motion. The operators $\mathcal{J}_2$, $\mathcal{J}_1$ respectively given by \eqref{J2}, \eqref{J1} correspond (up to a constant and a gauge transformation) to the symmetries of the Dunkl oscillator in the plane obtained in \cite{Genest-2013-04} which were found to generate the Schwinger-Dunkl algebra $sd(2)$ (see also \cite{Genest-2013-09} for the representation theory of $sd(2)$).
\section{Conclusion}
Recapping, we have shown that the model defined by the Hamiltonian \eqref{Hamiltonian-S2} on the two-sphere is superintegrable and exactly solvable. The constants of motion were explicitly obtained and were seen to be related to the separability of the Schr\"odinger equation associated to \eqref{Hamiltonian-S2} in different spherical coordinate systems. Moreover, it was observed that these symmetries generate a central extension of the Bannai--Ito algebra. The relation between the superintegrable system \eqref{Hamiltonian-S2} and the Racah problem for the $sl_{-1}(2)$ was also established. Furthermore, the contraction from the two-sphere to the Euclidean plane was examined and it was shown how the Dunkl oscillator in the plane and its symmetry algebra can be recovered in this limit.  

The results of this paper are complementary to those presented in  \cite{Genest-2013-12} where the connection between the combination of three $sl_{-1}(2)$ representations and the Dunkl Laplacian operator is used to study the $6j$ problem for $sl_{-1}(2)$ in the position representation and to provide a further characterization of the Bannai--Ito polynomials as interbasis expansion coefficients. Basis functions for irreducible representations of the Bannai--Ito algebra are also constructed in \cite{Genest-2013-12} and expressed in terms of the Dunkl spherical harmonics.

In view of the results obtained in \cite{Kalnins-2011-05} relating the generic model on the three-sphere to the bivariate Wilson polynomials, it would of interest to consider the analogous $S^3$ model with reflections. This could provide a framework for the study of multivariate $-1$ orthogonal polynomials. Also of interest is the study of the Dunkl oscillator models involving more complicated reflection groups.
\section*{Acknowledgments}
V.X.G. holds an Alexander-Graham-Bell fellowship from the Natural Sciences and Engineering Research Council of Canada (NSERC). The research of L.V. is supported in part by NSERC.
\section*{References}

\begin{thebibliography}{10}

\bibitem{Arfken-2012}
G.~B. Arfken, H.~Weber, and F.~E. Harris.
\newblock {\em {Mathematical Methods for Physicists}}.
\newblock Academic Press, 7 edition, 2012.

\bibitem{Arik-2003}
M.~Arik and U.~Kayserilioglu.
\newblock The anticommutator spin algebra, its representations and quantum
  group invariance.
\newblock {\em International Journal of Modern Physics A}, 18:5039--5046, 2003.

\bibitem{Bannai-1984}
E.~Bannai and T.~Ito.
\newblock {\em {Algebraic Combinatorics I: Association Schemes}}.
\newblock Benjamin/Cummings, 1984.

\bibitem{Dunkl-1989-01}
C.~F. Dunkl.
\newblock {Differential-difference operators associated to reflection groups}.
\newblock {\em Trans. Amer. Math. Soc.}, 311:167--183, 1989.

\bibitem{Dunkl-2001}
C.~F. Dunkl and Y.~Xu.
\newblock {\em {Orthogonal Polynomials of Several Variables}}.
\newblock Cambridge University Press, 2001.

\bibitem{Genest-2013-10}
V.~X. Genest, J-M. Lemay, L.~Vinet, and A.~Zhedanov.
\newblock {The Hahn superalgebra and supersymmetric Dunkl oscillator models}.
\newblock {\em J. Phys. A: Math. Theor.}, 46:505204, 2013.

\bibitem{Genest-2013-12}
V.~X. Genest, L.~Vinet, and A.~Zhedanov.
\newblock {A Laplace-Dunkl equation on $S^2$ and the Bannai--Ito algebra}.
\newblock {\em ArXiv:1312:6604}, 2013.

\bibitem{Genest-2013-12-1}
V.~X. Genest, L.~Vinet, and A.~Zhedanov.
\newblock {The Dunkl oscillator in three dimensions}.
\newblock {\em ArXiv:1312.3877}, 2013.

\bibitem{Genest-2013-tmp-2}
V.~X. Genest, L.~Vinet, and A.~Zhedanov.
\newblock {The Racah algebra and superintegrable models}.
\newblock {\em ArXiv:1312:3874}, 2013.

\bibitem{Genest-2013-09-02}
V.~X. Genest, L.~Vinet, and A.~Zhedanov.
\newblock { A "continuous" limit of the Complementary Bannai-Ito polynomials:
  Chihara polynomials}.
\newblock {\em SIGMA}, 10:38--55, 2014.

\bibitem{Genest-2013-04}
V.X. Genest, M.E.H. Ismail, L.~Vinet, and A.~Zhedanov.
\newblock {The Dunkl oscillator in the plane: I. Superintegrability, separated
  wavefunctions and overlap coefficients}.
\newblock {\em J. Phys. A: Math. Theor.}, 46:145201, 2013.

\bibitem{Genest-2013-09}
V.X. Genest, M.E.H. Ismail, L.~Vinet, and A.~Zhedanov.
\newblock {The Dunkl oscillator in the plane: II. Representations of the
  symmetry algebra}.
\newblock {\em Comm. Math. Phys. (to appear)}, 2013.

\bibitem{Genest-2013-02-1}
V.X. Genest, L.~Vinet, and A.~Zhedanov.
\newblock {Bispectrality of the Complementary Bannai--Ito polynomials}.
\newblock {\em SIGMA}, 9:18--37, 2013.

\bibitem{Genest-2013-tmp-1}
V.X. Genest, L.~Vinet, and A.~Zhedanov.
\newblock {Superintegrability in two dimensions and the Racah-Wilson algebra.}
\newblock {\em ArXiv:1307.5539}, 2013.

\bibitem{Genest-2013-07}
V.X. Genest, L.~Vinet, and A.~Zhedanov.
\newblock {The singular and 2:1 anisotropic Dunkl oscillators in the plane}.
\newblock {\em J. Phys. A: Math. Theor.}, 46:325201, 2013.

\bibitem{Genest-2012}
V.X. Genest, L.~Vinet, and A.~Zhedanov.
\newblock {The Bannai-Ito polynomials as Racah coefficients of the $sl_{-1}(2)$
  algebra}.
\newblock {\em Proc. Am. Soc.}, 142:1545--1560, 2014.

\bibitem{Gorodnii-1984}
M.F. Gorodnii and G.B. Podkolzin.
\newblock Irreducible representations of a graded {L}ie algebra.
\newblock In Yu.~M. Berezanskii, editor, {\em Spectral Theory of Operators and
  Infinite-Dimensional Analysis [in Russian]}. Inst. Matematiki Akad. Nauk
  Ukrain, 1984.

\bibitem{Jafarov-2011-05}
E.~I. Jafarov, N.~I. Stoilova, and J.~Van der Jeugt.
\newblock {Finite oscillator models: the Hahn oscillator}.
\newblock {\em J. Phys. A: Math. Theor.}, 44:265203, 2011.

\bibitem{Kalnins-2011-05}
E.~G. Kalnins, W.~Miller, and S.~Post.
\newblock {Two-variable Wilson polynomials and the generic superintegrable
  system on the 3-sphere}.
\newblock {\em SIGMA}, 7:51--76, 2011.

\bibitem{Kalnins-2013-05}
E.G. Kalnins, W.~Miller, and S.~Post.
\newblock {Contractions of 2D 2\textsuperscript{nd} order quantum
  superintegrable systems and the Askey scheme for hypergeometric orthogonal
  polynomials}.
\newblock {\em SIGMA}, 9:57--84, 2013.

\bibitem{Plyushchay-2001}
S.~M. Klishevich, M.~S. Plyushchay, and M.~R. de~Traubenberg.
\newblock {Fractional helicity, Lorentz symmetry breaking, compactification and
  anyons}.
\newblock {\em Nuclear Physics B}, 616(3):419--436, 2001.

\bibitem{Koekoek-2010}
R.~Koekoek, P.A. Lesky, and R.F. Swarttouw.
\newblock {\em {Hypergeometric orthogonal polynomials and their
  $q$-analogues}}.
\newblock Springer, 1\textsuperscript{st} edition, 2010.

\bibitem{Vinet-1996}
L.~Lapointe and L.~Vinet.
\newblock {Exact operator solution of the Calogero-Sutherland model}.
\newblock {\em Comm. Math. Phys.}, 178:425--452, 1996.

\bibitem{Macdonald-1999}
I.~G. Macdonald.
\newblock {\em {Symmetric functions and Hall polynomials}}.
\newblock Oxford University Press, 1999.

\bibitem{Miller-2013-10}
W.~Miller, S.~Post, and P.~Winternitz.
\newblock {Classical and quantum superintegrability with applications}.
\newblock {\em J. Phys. A: Math. Theor.}, 46:423001, 2013.

\bibitem{Kamefuchi-1982}
Y.~Ohnuki and S.~Kamefuchi.
\newblock {\em {Quantum Field Theory and Parastatistics}}.
\newblock Springer-Verlag, 1982.

\bibitem{Post-2011-12}
S.~Post, L.~Vinet, and A.~Zhedanov.
\newblock {An infinite family of superintegrable Hamiltonians with reflection
  in the plane}.
\newblock {\em J. Phys. A: Math. Theor.}, 44:505201, 2011.

\bibitem{Post-2011-10}
S.~Post, L.~Vinet, and A.~Zhedanov.
\newblock {Supersymmetric Quantum Mechanics with reflections}.
\newblock {\em J. Phys. A: Math. Theor.}, 44:435301, 2011.

\bibitem{Tsujimoto-2011-10}
S.~Tsujimoto, L.~Vinet, and A.~Zhedanov.
\newblock {From $sl\under{q}(2)$ to a parabosonic Hopf algebra}.
\newblock {\em SIGMA}, 7:93--105, 2011.

\bibitem{Vinet-2011}
S.~Tsujimoto, L.~Vinet, and A.~Zhedanov.
\newblock {Jordan algebra and orthogonal polynomials}.
\newblock {\em J. Math. Phys.}, 52:103512, 2011.

\bibitem{Tsujimoto-2012-03}
S.~Tsujimoto, L.~Vinet, and A.~Zhedanov.
\newblock {Dunkl shift operators and Bannai--Ito polynomials}.
\newblock {\em Adv. Math.}, 229:2123--2158, 2012.

\bibitem{Tsujimoto-2013-03-01}
S.~Tsujimoto, L.~Vinet, and A.~Zhedanov.
\newblock {Dual $-1$ Hahn polynomials: ``Classical'' polynomials beyond the
  Leonard duality}.
\newblock {\em Proc. Amer. Math. Soc.}, 141:959--970, 2013.

\bibitem{Vinet-2011-02}
L.~Vinet and A.~Zhedanov.
\newblock {A Bochner Theorem for Dunkl polynomials}.
\newblock {\em SIGMA}, 7:20--28, 2011.

\bibitem{Vinet-2011-01}
L.~Vinet and A.~Zhedanov.
\newblock {A `missing' family of classical orthogonal polynomials}.
\newblock {\em J. Phys. A: Math. Theor.}, 44:085201, 2011.

\bibitem{Vinet-2012-05}
L.~Vinet and A.~Zhedanov.
\newblock {A limit $q=-1$ for the Big $q$-Jacobi polynomials}.
\newblock {\em Trans. Amer. Math. Soc.}, 364:5491--5507, 2012.

\bibitem{Zhedanov-1991-11}
A.~Zhedanov.
\newblock {``Hidden symmetry'' of Askey-Wilson polynomials}.
\newblock {\em Theor. Math. Phys.}, 89:1146--1157, 1991.

\end{thebibliography}

\end{document}